\documentclass[%
 reprint,
 amsmath,amssymb,
 aps,
]{revtex4-2}

\usepackage{graphicx}
\usepackage{dcolumn}
\usepackage{bm}
\usepackage{hyperref}
\usepackage{xcolor}

\begin{document}

\preprint{APS/123-QED}

\title{First constraint on Born-rule violations at \\ high-energy colliders}

\author{Antony Valentini}
\email{a.valentini@imperial.ac.uk}
\affiliation{%
Abdus Salam Centre for Theoretical Physics, Imperial College London,
Prince Consort Road, London SW7 2AZ, United Kingdom
}%

\author{Mira Varma}
\email{mira.varma@yale.edu}
\affiliation{%
Department of Physics, Yale University, New Haven, CT 06520, USA
}%



\begin{abstract}
We obtain an experimental constraint on possible Born-rule violations at
high-energy colliders. We model Born-rule violations with differential
scattering cross sections $d\sigma/d\Omega$ subject to an angular smearing
by a narrow Gaussian of width $\varepsilon$ (with respect to $x=\cos\theta$
for scattering angle $\theta$). For large-angle Bhabha ($e^{+}e^{-}%
\rightarrow e^{+}e^{-}$) scattering, at a centre-of-mass energy $\sqrt{s}=29\,%
\mathrm{GeV}$, data from the PEP collider at SLAC allow us to set an upper
bound of $\varepsilon<0.042$ at $95\%$ confidence. This corresponds to a
Gaussian smearing over an angular range of twice the experimental bin
width, and hence provides a physically meaningful limit on deviations from
the Born rule. Future prospects for improving this limit are discussed.
\end{abstract}

\maketitle

\preprint{APS/123-QED}


\affiliation{Abdus Salam Centre for Theoretical Physics, Imperial College London,
Prince Consort Road, London SW7 2AZ, United Kingdom
}

\affiliation{Department of Physics, Yale University, New Haven, CT 06520, USA
}





\textit{Introduction---}In recent years, precision tests of quantum mechanics
have been extended into the high-energy domain by repurposing data from
colliders, providing evidence in particular for the key quantum phenomena of
entanglement and nonlocality in this regime~\cite{Barr24}. The Born
probability rule is another fundamental principle of quantum mechanics,
which can also be subject to precision tests at high energies~\cite{AnnFond}. 
It has been suggested that neutrino oscillations provide an effective
probe of novel triple-path interference effects~\cite{Minic22}, and that high-energy 
spin or polarisation probabilities can be deployed for targeted tests of the
Born rule in short-timescale processes~\cite{VV1}. In this
paper, we use data for large-angle Bhabha ($e^{+}e^{-}\rightarrow
e^{+}e^{-}$) scattering~\cite{Derr86}, to obtain an upper limit on Born-rule
violations at a centre-of-mass collision energy $\sqrt{s}=29\,\mathrm{GeV}$.

We model Born-rule violations as an angular smearing of the differential
scattering cross section. This is implemented by a narrow Gaussian of
standard deviation $\varepsilon $, defined with respect to $x=\cos \theta $
where $\theta $ is the scattering angle, and with a full width at half
maximum (FWHM) of $\simeq 2.4\varepsilon $. Because the collision data are
reported in angular bins of finite width $\Delta x=0.05$, there is an
intrinsic experimental smearing over the scale $\Delta x$, below which a
smearing from new physics would be undetectable. Our results yield an upper
bound $\varepsilon <0.042$ at $95\%$ confidence, excluding Gaussian smearing
with a FWHM $>2\Delta x$, thereby placing a meaningful constraint at
detectable scales. We also consider how this result might be improved.

While a detailed model of Born-rule violations at high energies will not be
needed here, it is useful to consider how such violations might arise. In
standard quantum mechanics, the Born rule is regarded as a basic law or
postulate. In contrast, in the de Broglie-Bohm pilot-wave approach~\cite%
{deB28,BV09,B52a,B52b,Holl93,EMP,AV25,OREP}---where a system with
configuration-space wave function $\psi (q,t)$ has a definite trajectory $%
q(t)$ (whose velocity is determined by $\psi $)---the Born rule $\rho
(q,t)=\left\vert \psi (q,t)\right\vert ^{2}$ describes a state of
statistical `quantum equilibrium' (analogous
to thermal equilibrium in classical physics) arising by dynamical relaxation~\cite%
{AV91a,VW05,TRV12,ACV14,AV20}. In principle, nonequilibrium distributions $%
\rho (q,t)\neq \left\vert \psi (q,t)\right\vert ^{2}$ are possible, yielding
Born-rule violations~\cite{AV91b,AV92,AV96,AV02a,AV09,EMP,AV25,OREP}. The
same logic applies to any deterministic hidden-variables theory: a
particular distribution $\rho _{\mathrm{QT}}(\lambda )$ of hidden variables $%
\lambda $ yields agreement with the Born rule for quantum measurements over
ensembles, while more general distributions $\rho (\lambda )\neq \rho _{%
\mathrm{QT}}(\lambda )$ break the Born rule~\cite{AV02b,AV04,AV07}, as has
recently been explored for high-energy spin and polarisation measurements~%
\cite{VV1}.

Born-rule violations in pilot-wave theory have been explored as a possible
explanation for reported large-scale anomalies in the cosmic microwave
background~\cite{AV07,AV10,CV13,CV15,VPV19}. A gravitational instability of
the Born rule may affect Hawking radiation from primordial black holes~\cite{AV23}, 
in part motivating an experimental test of the Born rule in space~\cite{Ahm24}. 
Finally, the de Broglie velocity equation for trajectories $q(t)$ requires 
regularisation at nodes ($\psi =0$), which can be achieved by smearing the velocity 
field in configuration space, resulting in a smearing of the equilibrium Born 
rule~\cite{AnnFond}. This last possibility has motivated a search for Born-rule 
violations in short-timescale processes at colliders~\cite{VV1}.

In this paper, we take a phenomenological approach. In high-energy physics,
the Born rule is applied to calculate probabilities $\left\vert \left\langle
f\right\vert \hat{S}\left\vert i\right\rangle \right\vert ^{2}$ for
transitions from initial states $\left\vert i\right\rangle $ to final states 
$\left\vert f\right\rangle $, where $\left\langle f\right\vert \hat{S}%
\left\vert i\right\rangle $ are $S$-matrix elements. The resulting
differential scattering cross sections take the form
\begin{equation}
\frac{d\sigma }{d\Omega }\propto \left\vert \mathcal{M}\right\vert ^{2}\,,
\end{equation}%
where $\mathcal{M}$ is the Feynman amplitude. While any measurement of a cross 
section is in a sense a test of the Born rule, of interest here are targeted tests 
aimed at setting precise limits on possible Born-rule violations. To this end, we 
may consider smeared cross sections%
\begin{equation}
\left( \frac{d\sigma }{d\Omega }\right) _{\mathrm{smear}}=\int d\Omega
^{\prime }\,\delta _{\varepsilon }(\Omega ^{\prime }-\Omega )\left( \frac{%
d\sigma }{d\Omega ^{\prime }}\right) \,,  \label{smear}
\end{equation}
where symbolically $\Omega $ represents a point $(\theta ,\phi )$ on the unit sphere, 
and $\delta _{\varepsilon }(\Omega ^{\prime }-\Omega )$ is a normalised narrow 
Gaussian of width $\varepsilon $. This amounts to a smearing of  the usual Born 
rule on angular scales $\varepsilon $. Our aim is to set an experimental limit on the 
value of $\varepsilon $.

If smeared cross sections of the form (\ref{smear}) were observed, on scales 
$\varepsilon $ larger than the experimental smearing from angular
binning of data, it could provide evidence for new physics, possibly along
the lines indicated. For our purposes, setting experimental limits on $%
\varepsilon $ with available data allows us to set useful, model-independent
constraints on the Born rule in the high-energy regime.

\textit{Bhabha scattering with a smeared Born rule---}We consider high-energy
Bhabha ($e^{+}e^{-}\rightarrow e^{+}e^{-}$) scattering in the centre-of-mass
frame, which in our case coincides with the laboratory frame. This process
is particularly convenient for testing the Born rule because its cross
section is large, the resulting statistical uncertainties are small, and the
Standard-Model prediction can be calculated accurately. Bhabha scattering
has recently been considered in a study of entanglement distribution~\cite{Blas24}, 
and as a means to test entanglement for freely travelling 
electron-positron pairs~\cite{Gao25}.

At a centre-of-mass energy $\sqrt{s}=29\,\mathrm{GeV}$, the process is
essentially governed by QED with only small electroweak corrections. To
lowest order, including both pure-QED and electroweak effects, the standard
differential scattering cross section is given by~\cite{Derr86} 
\begin{equation}
\begin{split}
\frac{d\sigma}{d\Omega} &=\frac{\alpha^{2}}{2s} \Biggl[ |A_{1}|^{2}\left(%
\frac{s}{t}\right)^{2} +|A_{2}|^{2}\left(\frac{t}{s}\right)^{2} \\
&\qquad +\frac{1}{2} \left(|A_{3}|^{2}+|A_{4}|^{2}\right) \left(1+\frac{t}{s}%
\right)^{2} \Biggr],
\end{split}
\label{diff_cross_sec_lowest_order}
\end{equation}
where $t=-s(1-\cos\theta)/2$ with scattering angle $\theta$. For energies $%
\sqrt{s}$ much lower than the $Z$-boson mass $M_{Z}$, the
electroweak amplitudes are given by

\begin{align*}
A_{1}& =1+(g_{\mathrm{V}}^{2}-g_{\mathrm{A}}^{2})\chi (t)\,, \\
A_{2}& =1+(g_{\mathrm{V}}^{2}-g_{\mathrm{A}}^{2})\chi (s)\,, \\
A_{3}& =1+\frac{s}{t}+(g_{\mathrm{V}}-g_{\mathrm{A}})^{2}\left[ \chi (s)+%
\frac{s}{t}\chi (t)\right] \,, \\
A_{4}& =1+\frac{s}{t}+(g_{\mathrm{V}}+g_{\mathrm{A}})^{2}\left[ \chi (s)+%
\frac{s}{t}\chi (t)\right] \,,
\end{align*}
where 
\[
\chi (q^{2})=\frac{G_{\mathrm{F}}}{\pi \alpha \sqrt{8}}\frac{q^{2}M_{Z}^{2}}{(q^{2}-M_{Z}^{2})}\,.
\]%
Here $q$ is the usual 4-momentum transfer, $\alpha $ and $G_{\mathrm{F}}$
are respectively the electromagnetic and Fermi couplings, while $g_{\mathrm{V%
}}$ and $g_{\mathrm{A}}$ are respectively the vector and axial-vector
couplings.

In terms of $x=\cos \theta $, we have $d\Omega =dxd\phi $. Since $d\sigma
/d\Omega $ is independent of the azimuthal angle $\phi $, we may define $d\sigma /dx=2\pi d\sigma /d\Omega$. 
As a phenomenological construct, the standard $d\sigma /dx$ may be replaced by a smeared expression
\begin{equation}
\left( \frac{d\sigma }{dx}\right) _{\mathrm{smear}}=\int_{x-a}^{x+a}dx^{%
\prime }\,\delta _{\varepsilon }(x^{\prime }-x)\,\left( \frac{d\sigma }{%
dx^{\prime }}\right) \,,  \label{smear_x}
\end{equation}
where $\delta _{\varepsilon }(x^{\prime }-x)$ is a normalised Gaussian of 
width $\varepsilon $ centred at $x^{\prime }=x$, and $a$ is a range parameter chosen to avoid 
edge effects at $x^\prime=\pm 1$ (as well as avoiding the divergence in $d\sigma /dx^{\prime }$
at $x^{\prime }=1$). We will study large-angle data in the range $-0.55<x<0.55$. 
We then require $a<0.45$, which also ensures that (\ref{smear_x}) is defined over the whole
data range.

The smeared differential cross section (\ref{smear_x}) will be fit to data
from Ref.~\cite{Derr86} (Table XII, column 7), to obtain a bound on $%
\varepsilon $. In Ref.~\cite{Derr86}, the same data were used to obtain
bounds on possible high-energy cutoff scales $\Lambda _{\pm }$.

As a check on robustness, our fitting procedure was repeated with $a=0.25$, $%
\,0.3$, $\,0.35$ and $0.4$. The best-fit $\varepsilon $ and $95\%$
confidence upper bound were both unchanged to four decimals. Our reported
results below use $a=0.4$. Our best-fit $\varepsilon =0.024$ corresponds to a
smearing Gaussian $\delta _{\varepsilon }(x^{\prime }-x)$ with a FWHM $%
\simeq 0.05$. This is less than one tenth of the integration range in (\ref%
{smear_x}) (with $a=0.4$). The tails of the Gaussian are then completely
negligible outside the integration range.

\textit{Available datasets---}We use large-angle Bhabha scattering data from
the High Resolution Spectrometer (HRS) at PEP, which provides bin-by-bin
measurements of the differential cross section at $\sqrt{s}=29\,\mathrm{GeV}$%
, as given in Table XII of Ref.~\cite{Derr86}. The measurements in the
fiducial region $|x|<0.55$ are reported in 22 bins of uniform width $\Delta
x=0.05$, with bin centres ranging from $x=-0.525$ to $0.525$. The angular
distribution is constructed using the scattering angles of both outgoing charged 
particles, with each track assigned a weight of $0.5$. Table XII of Ref.~\cite{Derr86} 
also reports the measured differential cross section with the $\mathcal{O}(\alpha ^{3})$ 
QED radiative contribution removed, the corresponding $\mathcal{O}(\alpha ^{2})$ 
electroweak-plus-QED prediction, event counts, acceptance factors, and QED and 
electroweak correction factors used in constructing these results.

To compare with the $\mathcal{O}(\alpha ^{2})$ Born-rule prediction (whether
smeared or unsmeared), we use data with the $\mathcal{O}(\alpha ^{3})$
effects removed. These corrected data are shown in Fig.~\ref%
{fig:bhabha_unsmeared_eq3}, together with the standard (unsmeared) $\mathcal{%
O}(\alpha ^{2})$ Born-rule prediction (\ref{diff_cross_sec_lowest_order}).

\begin{figure}[t]
\centering \includegraphics[width=0.7\columnwidth]{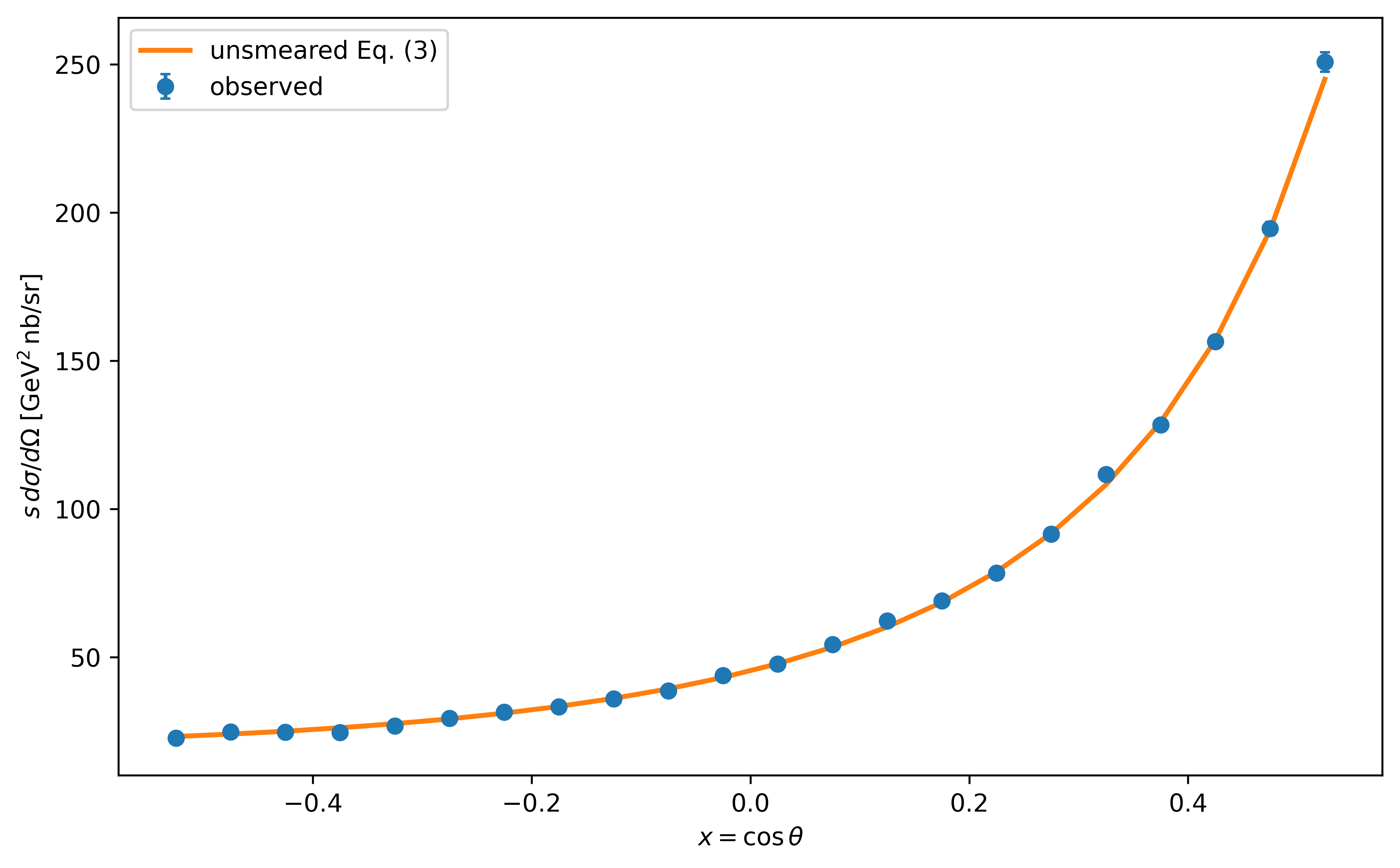}  
\caption{HRS Bhabha scattering data with $\mathcal{O}(\protect\alpha ^{3})$
effects removed (Table XII of Ref.~\protect\cite{Derr86}) compared with the
standard (unsmeared) $\mathcal{O}(\protect\alpha ^{2})$ Born-rule prediction
(\protect\ref{diff_cross_sec_lowest_order}).}
\label{fig:bhabha_unsmeared_eq3}
\end{figure}


We found it convenient to use a dataset from 1986, at a relatively low
energy, for two reasons. First, we require a dataset sufficiently
transparent to allow us to reconstruct, bin by bin, how the measured event
counts were transformed into the published angular observables. Second,
working at lower energies minimises the need to model and subtract
higher-energy radiative or electroweak corrections. These points may be
illustrated by briefly discussing some other available datasets.

For example, the 2007 ALEPH $e^{+}e^{-}\rightarrow e^{+}e^{-}$ analysis~\cite%
{ALEPH:2006jhv} tabulates the measured differential cross sections, but its
binned maximum-likelihood fit combines data from several centre-of-mass
energies. Since the precise multi-energy likelihood implementation is not
tabulated, reproducing their published result would require reconstructing the full fit. 
Similarly, the 2006 DELPHI $e^{+}e^{-}\rightarrow e^{+}e^{-}$ measurement~\cite{DELPHI:2005wxt} 
publishes bin-by-bin information, including selection efficiencies, backgrounds, and differential 
cross sections, making it a promising candidate for strengthening our bounds in future work. 
However, the published limits on short-distance deviations from QED are extracted using 
combined fits to cross sections and forward-backward asymmetries over all LEP II centre-of-mass 
energies, including QED radiative corrections and correlated systematic uncertainties. 
Reproducing these limits would require reconstructing the full multi-energy fit. For our first 
collider constraint on Born-rule violations, we use the single-energy dataset from 
Ref.~\cite{Derr86}, which allows a direct comparison with the smeared cross section using 
only the published tables.

Among the closest available alternatives to Ref.~\cite{Derr86} is the TASSO
1988 measurement~\cite{TASSO:1987cag}, which also published absolute Bhabha
differential cross sections. However, only six bins have $\Delta x=0.05$,
while the remaining thirteen have $\Delta x=0.10$. Since $\varepsilon $
directly characterises angular smearing, the relevant experimental scale is
the published bin width. TASSO therefore averages away angular structure
over most of the range at a scale about twice as large as the uniform $%
\Delta x=0.05$ bins of Ref.~\cite{Derr86}, making it less suitable for our
purposes.

There is a natural concern that, since the dataset we use is from 1986,
repeating the experiment with a modern detector might significantly improve
the precision. However, in Ref.~\cite{Derr86} the bin-by-bin Bhabha
differential cross section in the large-angle region is already measured at the 
percent level. In later LEP2 measurements, the precision is not much better and, 
in some cases, significantly worse. For example, in the ALEPH measurement~\cite{ALEPH:2006jhv} 
some bins are measured at the $1\%-3\%$ level, but others have uncertainties of 
$10\%-30\%$. The DELPHI measurement~\cite{DELPHI:2005wxt} shows a similar pattern: the most accurate bins are measured to a few percent, while several bins have uncertainties of around $30\%$. Since 
we wish to study the differential cross section in angular bins at a given energy, 
Ref.~\cite{Derr86} provides more percent-level measurements than the later LEP2 experiments.

Finally, although more recent experiments achieve a smaller detector angular
resolution, they often report in coarser bins, limiting the amount of
information one can obtain from the published results. For example, Ref.~%
\cite{DELPHI:2005wxt} reports a polar-angle precision of approximately 3.5
mrad for the HPC electromagnetic clusters, finer than the 6.5 mrad HRS
charged-track polar-angle detector resolution of Ref.~\cite{Derr86}.
However, the published DELPHI angular distributions for Bhabha scattering
are reported with bin widths $\Delta x=0.09$ or $0.18$ depending on the
angular region. This is substantially larger than the uniform bin width $%
\Delta x=0.05$ of Ref.~\cite{Derr86}.

\begin{figure}[t]
\centering  \includegraphics[width=0.7\columnwidth]{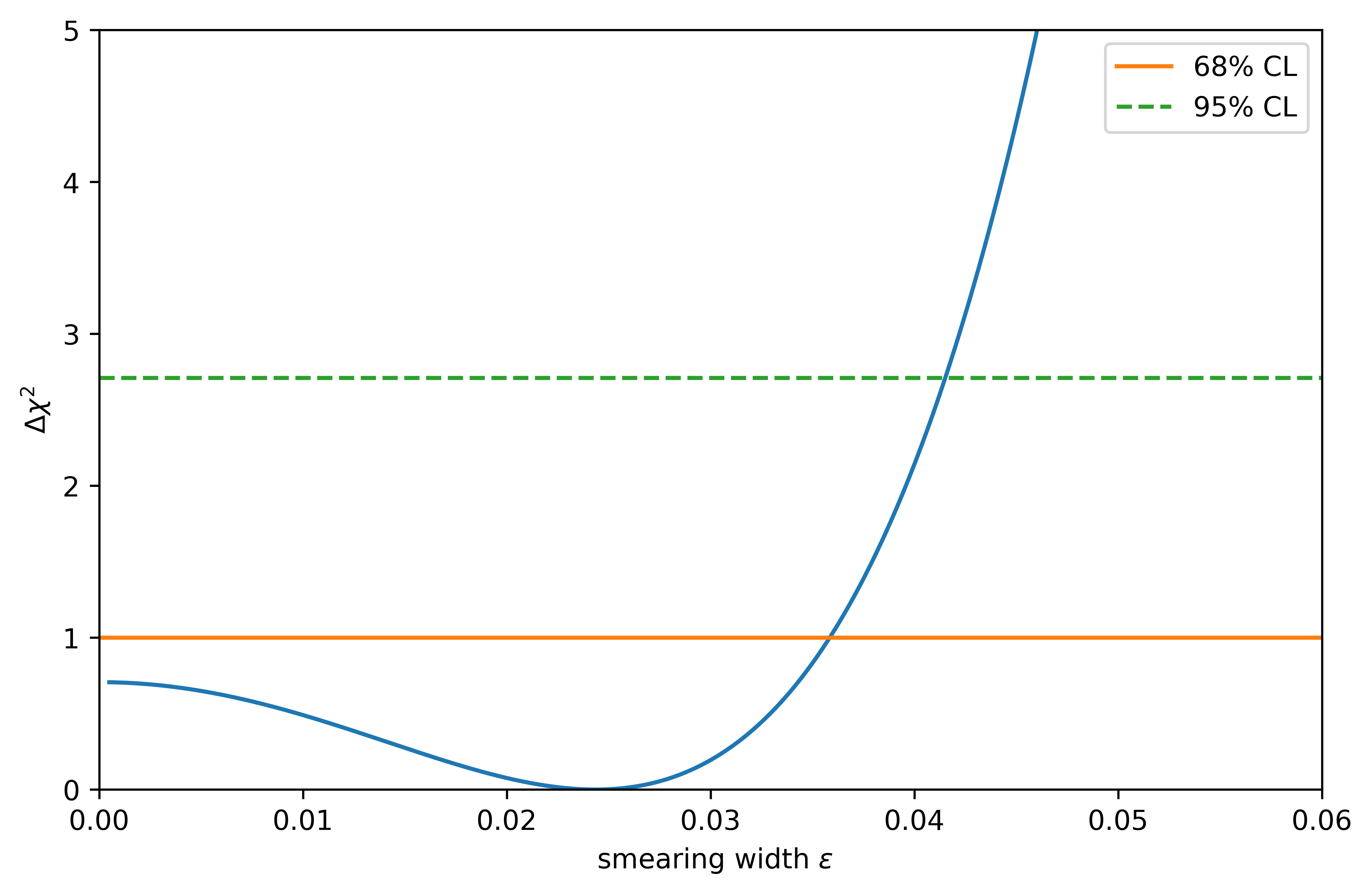}  
\caption{$\Delta \protect\chi ^{2}$ scan for the smeared expression (\protect
\ref{smear_x}) (with $a=0.4$). The best fit occurs at $\protect\varepsilon %
=0.024$. The full orange and dashed green lines respectively represent the
one-sided $68\%$ and $95\%$ confidence limits.}
\label{fig:bhabha_dchi2_scan}
\end{figure}

\begin{figure}[t]
\centering   
\includegraphics[width=0.7\columnwidth]{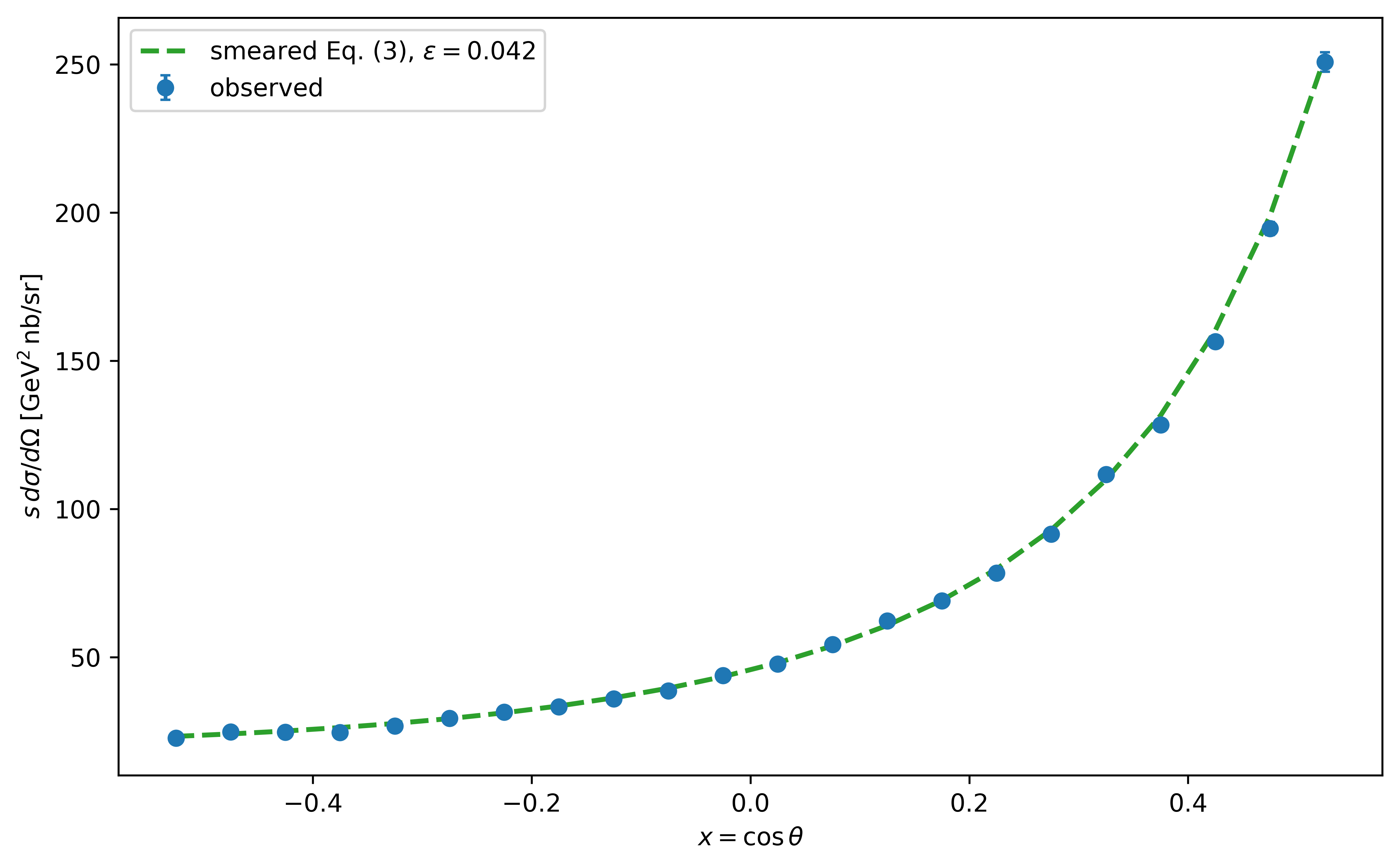}  
\caption{HRS Bhabha scattering data with $\mathcal{O}(\protect\alpha ^{3})$
effects removed (Table XII of Ref.~\protect\cite{Derr86}) compared with the $%
\mathcal{O}(\protect\alpha ^{2})$ smeared prediction (\protect\ref{smear_x})
at the $95\%$ confidence limit with $\protect\varepsilon =0.042$.}
\label{fig:bhabha_smeared_95}
\end{figure}

\textit{Likelihood scan and result---}Our analysis performs a likelihood scan as a function 
of the Born-rule smearing parameter $\varepsilon $. We use $N_{\mathrm{bins}}=22$ angular bins from Table XII of Ref.~\cite{Derr86}. For each $\varepsilon $, (\ref{smear_x}) is 
used to calculate the corresponding smeared differential cross section, from which we compute 
the probability for the observed data (that is, the likelihood for $\varepsilon$ given the 
data). Repeating over a range of $\varepsilon $ values gives the likelihood profile, 
from which we construct $\Delta \chi ^{2}$ as a function of $\varepsilon$ (Fig.~\ref{fig:bhabha_dchi2_scan}). The confidence intervals are obtained using Wilks' theorem. By construction, $\Delta \chi_{\min}^{2}=0$.

Due to uncertainties in the luminosity and the $\mathcal{O}(\alpha^3)$ QED calculation, the 
theoretical prediction in each bin is multiplied by a dimensionless nuisance parameter $A$, which is constrained by a Gaussian distribution centred at $A=1$ with width $\sigma_A=0.0117$ (obtained by adding in quadrature the quoted $0.6\%$ luminosity uncertainty and the estimated $1\%$ systematic uncertainty arising from the $\mathcal{O}(\alpha^3)$ QED calculation). For each fixed value of $\varepsilon$ (scanned over a range $0\leq\varepsilon\leq0.2$), we profile over $A$ by minimising the full $\chi^2$ including the Gaussian constraint. This defines the one-dimensional profile $\chi^2_{\rm prof}(\varepsilon)=\min_A\chi^2(\varepsilon,A)$, which is then used to construct the $\Delta\chi^2$ curve in Fig.~\ref{fig:bhabha_dchi2_scan}. 

The best-fit smearing width is found to be 
\begin{equation}
\varepsilon =0.024 
\end{equation}%
(with best-fit $A = 0.99$), where $\chi_{\min }^{2}=22.3$, yielding a 
good fit with $\chi_{\min}^{2}/n_{\mathrm{dof}}\simeq 1.06$ (for $n_{\mathrm{dof}}=N_{\mathrm{bins}}+1 -2=21$ degrees of freedom). This is nearly half the experimental angular
bin width, $\Delta x/2=0.025$. The FWHM ($\simeq 2.4\varepsilon =0.058$) of
the best-fit smearing Gaussian closely matches the bin width $\Delta x=0.05$. Since the data 
are averaged over bins, if the theoretical curve is
averaged over a comparable length scale, we indeed expect to obtain a slight
improvement to the fit with no need for new physics.

Our main result is the upper bound 
\begin{equation}
\varepsilon <0.042
\end{equation}%
at $95\%$ confidence (Fig.~\ref{fig:bhabha_dchi2_scan}). This puts a
meaningful constraint on any anomalous smearing of the Born-rule prediction.
The Gaussian FWHM at the $95\%$ limit is almost exactly twice the bin
width, hence smearing at this scale would correspond to new physics and not
be related to a simple binning of the data. In other words, smearing widths $%
\varepsilon $ comparable to or larger than the intrinsic experimental
angular bin width $\Delta x$ are inconsistent with the data. This result is
physically significant, and we rule out smearing widths $\varepsilon >0.042$
at $95\%$ confidence.

Fig.~\ref{fig:bhabha_smeared_95} compares the Bhabha scattering data from
Ref.~\cite{Derr86} with the maximally smeared prediction (\ref{smear_x}) at $%
95\%$ confidence. Even at this limiting value, (\ref{smear_x}) remains close
to the data over the full angular range, departing slightly towards the
right.

\textit{Future prospects for improved limits---}The smallest smearing scale
that can be meaningfully probed experimentally is given by the effective
angular resolution of the published measurement. For data in Ref.~\cite%
{Derr86}, this is set by the bin width $\Delta x=0.05$, since the detector
angular resolution is substantially smaller. In
principle, $\Delta x$ is an analysis choice. However, in practice, if $%
\Delta x$ is too small, the measured differential cross sections will be
dominated by statistical errors. There is an unavoidable trade-off between
probing at smaller angular scales and obtaining more accurate results. For
the data in Ref.~\cite{Derr86}, the quoted bin uncertainties are mainly
statistical. The HRS charged-track polar-angle resolution is 6.5 mrad,
corresponding to $\delta x= 0.0065$ at $x=0$, an order of magnitude
smaller than the bin width. Thus, for the present study, the bin width is
more relevant than the raw detector resolution.

For a total number $N$ of events, with $n_{i}$ events assigned to the $i$th
bin, we have an estimated probability $p_{i}\approx n_{i}/N$ with a relative
error $\Delta p_{i}/p_{i}\approx 1/\sqrt{n_{i}}$ (where $p_{i}\propto
(d\sigma /d\Omega )_{i}$). A fixed relative error requires us to fix $%
n_{i}\approx Np_{i}\approx N\rho _{i}\Delta x$ (where $\rho _{i}$ is the
theoretical probability density at the centre of the $i$th bin), hence the
bin width scales approximately as $\Delta x\propto 1/N$. Increasing the
luminosity $L$ over a given time then allows us to decrease the bin width $%
\Delta x\propto 1/L$, while maintaining a fixed statistical uncertainty per
bin. However, once $\Delta x$ becomes comparable to or smaller than $\delta x
$, a further increase in $L$ will not improve the effective angular
resolution unless $\delta x$ is also reduced.

This point is relevant when comparing Ref.~\cite{Derr86} with more
modern measurements. The HRS analysis employed $84,\!423$ observed Bhabha events. In
contrast, the 2024 BESIII large-angle Bhabha luminosity analysis of Ref.~%
\cite{BESIII:2024lbn} reports $~4-7\times 10^{8}$ observed events in each of
its three 2021--2024 data sets. This alone would allow the bin width to be
reduced by about three orders of magnitude without affecting the statistical
errors, if only the detector angular resolution could also be reduced by a
comparable factor. Should such improvements in detector resolution be
forthcoming, modern data would allow a substantially stronger bound than the
one obtained here, provided the data were
published with sufficiently fine and uniform angular bins.

At colliders, the polar-angle resolution $\delta \theta =\sqrt{(\Delta
\theta _{\mathrm{res}})^{2}+(\Delta \theta _{\mathrm{ms}})^{2}}$ is
typically limited by the track measurement resolution $\Delta \theta _{%
\mathrm{res}}$ and an error $\Delta \theta _{\mathrm{ms}}$ from multiple
scattering within the detector material. Ref.~\cite{Drasal:2018zij} provides
analytic expressions for $\Delta \theta _{\mathrm{res}}$, $\Delta \theta _{%
\mathrm{ms}}$ for a standard solenoid spectrometer with a constant magnetic
field and $N+1$ equidistant detector planes,%
\begin{equation}
\Delta \theta _{\mathrm{res}}\approx \frac{2\sigma _{z}\sin ^{2}\theta }{%
L_{0}}\sqrt{\frac{3}{N+3}}
\end{equation}%
(assuming $2/N<<N+3$), and 
\begin{equation}
\Delta \theta _{\mathrm{ms}}\approx \frac{0.0136~\mathrm{GeV}/c}{\beta p_{T}}%
\sqrt{\frac{d\sin \theta }{X_{0}}},
\end{equation}%
where $\sigma _{z}$ is the position measurement resolution in the $z$%
-direction, $L_{0}$ is the perpendicular distance through the detector planes, $d$ is the thickness of a single detector plane, and $X_{0}$ is the radiation length. As an example of orders of magnitude, from Ref.~\cite{Drasal:2018zij}, taking $L_{0}=1~\mathrm{m}$, $\sigma _{z}=50~\mu \mathrm{m}
$, $N+1=10$, $\beta \approx 1$ and $\theta =\pi /2$, we have $\delta \theta
\approx \Delta \theta _{\mathrm{res}}=0.05~\mathrm{mrad}$ (where $\Delta
\theta _{\mathrm{ms}}$ can be neglected for $p_{T}>50$ $\mathrm{GeV}/c$).

The question is whether $\delta \theta $ could be significantly reduced in a
forthcoming experiment. If we neglect $\Delta \theta _{\mathrm{ms}}$, we have
roughly 
\begin{equation}
\delta \theta \sim \frac{\sigma _{z}}{L_{0}\sqrt{N}}.
\end{equation}
To decrease $\delta \theta $ by a factor of 10, we could decrease $\sigma_{z}$ by a factor 
of 10, increase $L_{0}$ by a factor of 10, or increase $N$ by a factor of 100 (or some 
combination of improvements in $\sigma _{z}$, $L_{0}$ and $N$). However, significantly 
increasing $L_{0}$ or $N$ may be less practical than decreasing $\sigma_{z}$. This suggests 
aiming for better technology to improve the $z$-resolution.

In the ATLAS experiment~\cite{ATLAS:2010ylv}, the angular resolution of the inner detector 
varies over the range $\delta \theta =$ 0.5 mrad -- 1 mrad (depending on $\theta $), up to an order of magnitude better than the detector resolution 
$\delta \theta \approx 6.5$ mrad in Ref.~\cite{Derr86}. Similarly, in the CMS 
experiment~\cite{CMS:2009dvy}, the angular resolution of the silicon tracker is 
$\delta \theta \approx 0.5$ mrad, a full order of magnitude better than in Ref.~\cite{Derr86}. 
We also note that, in studies for the proposed Future Circular Collider~\cite{Blondel:2019ylv}, 
it is hoped to attain $\delta \theta \approx 0.1$ mrad, still less than two orders of magnitude 
better than in Ref.~\cite{Derr86}. Therefore, in present and forthcoming colliders, we can 
only hope for an improvement of less than two orders of magnitude over the limits found in 
this paper.

\textit{Conclusion---}We have obtained a limit on potential Born-rule
violations at colliders, using data for Bhabha scattering at $\sqrt{s}=29\,%
\mathrm{GeV}$, in the form of a limit on the angular smearing width $%
\varepsilon $, which we have bounded at $\varepsilon <0.042$ at $95\%$
confidence. It appears difficult to improve this result significantly with
data from current or forthcoming colliders. Remarkably, as it stands, the
Born rule is tested rather poorly by the standards of what might be expected
for a fundamental principle of physics. Significantly improving our result
may require new techniques tailored towards testing the Born rule in
high-energy physics.
\\ 

\textit{Acknowledgement---}M. V. is supported by the DOE Office of High Energy Physics under Grant No. DE-SC0017660.

\end{document}